\documentclass[epj]{svjour}
\usepackage{graphicx}
\begin{document}
\title{Hadron spectrum and hadrons in the nuclear medium}
\author{M. J. Vicente Vacas
}                     
%
%
\institute{Departamento de F\'{\i}sica Te\'orica and IFIC,
Centro Mixto Universidad de Valencia CSIC;
Institutos de Investigaci\'on de Paterna, Aptdo. 22085, 46071 Valencia, Spain.}
\date{Received: date / Revised version: date}
%
\abstract{Some recent developments in chiral dynamics of hadrons and hadrons in a medium 
are presented. Unitary schemes based on chiral Lagrangians describe some hadronic states as being dynamically generated
resonances.
 We discuss how standard quantum many body techniques can be used to calculate the properties of these dynamically generated and other hadrons in the nuclear medium. We present some results for
vector mesons ($\rho$ and $\phi$), scalar mesons ($\sigma$, $\kappa$, $a_0(980)$, $f_0(980)$), the
$\Lambda(1520)$ and for the in-medium baryon-baryon interaction. 
} 
\maketitle
\section{Introduction}
\label{intro}

Coupled channels unitary models based on the chiral Lagrangians have provided a framework that allows to extend the calculations to higher energies than standard $\chi$PT and have been very successful in the description of hadronic phenomenology. These models reproduce well meson-meson phase shifts up to energies above 1 GeV \cite{Oller:1997ti} and are able to generate a number of well established baryonic resonances  like $N^*(1535)$, $\Lambda(1405)$, $\Lambda(1670)$, $\Sigma(1620)$, $\Xi(1690)$ and others
\cite{Lutz:2001yb,Oset:2001cn,Garcia-Recio:2002td,Borasoy:2004kk,Inoue:2001ip}. These resonances appear dynamically as poles of the meson baryon scattering amplitudes. The analysis of the residues of the poles in the different scattering channels gives the couplings and branching ratios of the resonances.

 An important problem in nuclear physics is the understanding of the properties
of hadrons at finite baryonic densities. Many different theoretical approaches
to this problem can be found in the  literature.  A quite interesting question 
is if the very special nature of the dynamically generated resonances obtained
in chiral unitary models will have consequences for their in medium properties.
It looks clear that this description of these particles  has some implications.
For instance, the $\Lambda(1405)$ appears in these models as a resonant state in
$\bar{K} N$ and/or $\pi\Sigma$ scattering. The changes in the properties of the
constituents $\bar{K},\, N,\, \pi,\,\Sigma$ will also change the results for their
scattering amplitude and thus for its resonant states like the $\Lambda(1405)$.

The basic idea of the approach we present here is to start from these models
which describe adequately the properties of several  hadronic resonances in
vacuum and their couplings to pseudoscalar mesons. Then, we incorporate  
nuclear medium effects by including  vertex corrections and a suitable
modification of the pseudoscalar meson and the baryon propagators, which can be
done in terms of density dependent selfenergies. The selfenergies are studied by
means of a many body calculation  accounting for their interactions with the
nuclear medium.  With this formulation we automatically consider the presence of
additional decay channels which are not present in vacuum, since they require an
interaction with one or more nucleons in the medium. 

We will start discussing the $\pi$ and $K$ selfenergies. They experience a
sizable renormalization at finite densities as a consequence of their
interactions with the nucleons as it has been found in many theoretical
calculations and experimental results. Then, using these selfenergies as
ingredients, we study some medium effects on vector mesons ($\rho$ and $\phi$),
dynamically generated scalar mesons ($\sigma$, $\kappa$, $a_0(980)$ and
$f_0(980)$) and for the $\Lambda(1520)$ resonance. Finally, we will discuss
briefly some recent results on the $NN$ interaction in nuclear matter.

\section{Pion and Kaons selfenergies in nuclear matter}
\label{sec:pio} 

\begin{figure} \begin{center}
\resizebox{0.40\textwidth}{!}{%
\includegraphics{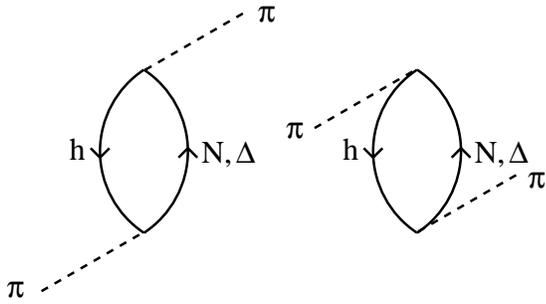} } 
\caption{Diagrams
contributing to $\pi$ selfenergy in the nuclear medium.} 
\label{fig:pion0}      
\end{center} 
\end{figure}

\begin{figure} \resizebox{0.45\textwidth}{!}{%
\includegraphics{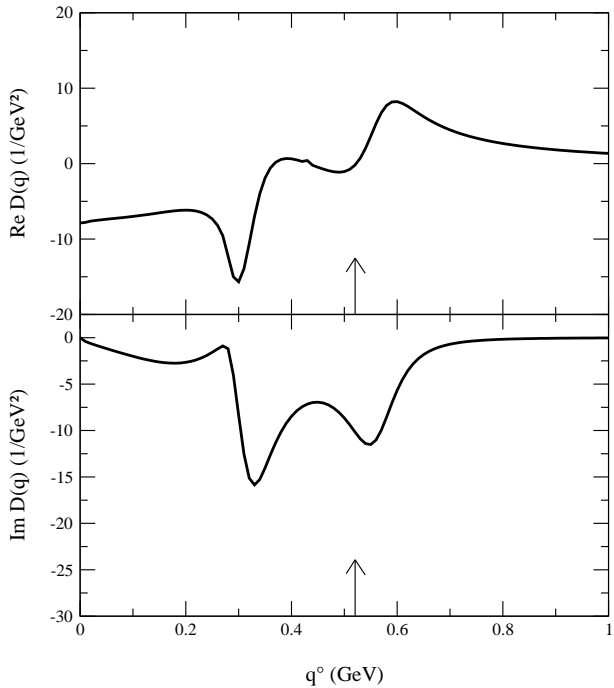} } 
\caption{ Real and imaginary part of the $\pi$
propagator at $\rho=\rho_0$ as a function of the energy for a pion with momentum
$q=500$ MeV. The arrow indicates the position of the pion pole in vacuum.}
\label{fig:pion1}       
\end{figure}

The pion selfenergy in the nuclear medium at intermediate energies is relatively
well known. It is mainly driven by its p-wave coupling to $p-h$ and $\Delta-h$
excitations as depicted in  Fig. \ref{fig:pion0}. 
Higher order terms obtained by
the iteration of these excitations, considering also short range correlations,
and the small contribution of s-wave scattering are also included in the
calculations. Analytical expressions, as well as results can be found in Refs.
\cite{ericson}. In Fig. \ref{fig:pion1} we show the results for the pion
propagator. 
Three wide peaks
appear in its imaginary part corresponding to $p-h$ excitation, the pion peak
that occurs at lower energies than in vacuum and $\Delta-h$ excitation
respectively. This model has been widely tested and agrees well with
pion-nucleus phenomenology from low to intermediate energies.

The kaon selfenergy is smooth at low energies and a simple $t\rho$ approximation can be used
\cite{Oset:2000eg}.  The situation is more controversial for the antikaons
with recent experimental data that might be interpreted as very deep kaonic atoms 
\cite{Suzuki:2004ep,Agnello:2005qj} suggesting very strong potentials, incompatible with the
calculations based on chiral models, although alternative interpretations of these results have been presented
\cite{Oset:2005sn}. See also  Ref. \cite{Mares:2006vk} and references therein.

The model used in what follows is based on  the chiral unitary model in coupled channels for the s-wave $\bar{K} N$ scattering of Ref. \cite{Oset:1998it}. This model is quite successful in the description of many observables, like
threshold ratios of $K^- p$ to inelastic channels and $K^- p$ cross sections in the elastic and inelastic channels. 
A selfconsistent calculation of the selfenergy in the nuclear medium  was later
 carried out in Refs.
\cite{Ramos:1999ku,Tolos:2006bj} and includes the p-wave excitation of $\Lambda h$, $\Sigma h$ and  $\Sigma(1385) h$.
The results of these works show an antikaon spectral function quite wide at normal nuclear densities and with the peak displaced toward low energies. The optical potential from Ref. \cite{Ramos:1999ku} also compares well with kaonic atoms data \cite{Baca:2000ic} although other  phenomenological potentials with a much deeper attraction also show a good agreement \cite{Cieply:2001yg}.

\section{Vector mesons ($\rho$ and $\phi$)}
\label{sec:vec}
 
In medium properties of the $\rho$ meson  have been extensively studied 
\cite{Hatsuda:1992ez,Leupold:1998dg,Klingl:1997kf,Mallik:2001gv,Asakawa:1992ht,Chanfray:1993ue,Herrmann:1993za,Peters:1998va,Lutz:1999jn} after some predictions of scaling laws implying strong
droppings of the $\rho$ meson mass even at normal nuclear density. On the other hand, this mass reduction could be observed through the vector meson leptonic decays, free from strong final state distortions. Here, we will briefly present our approach to this topic.

\begin{figure}
\resizebox{0.45\textwidth}{!}{%
  \includegraphics{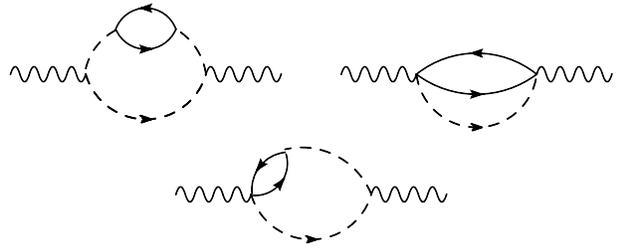}
}
\caption{Diagrams contributing to $\rho$ selfenergy in the nuclear medium.}
\label{fig:rho1}       
\end{figure}

\begin{figure}
\resizebox{0.45\textwidth}{!}{%
  \includegraphics{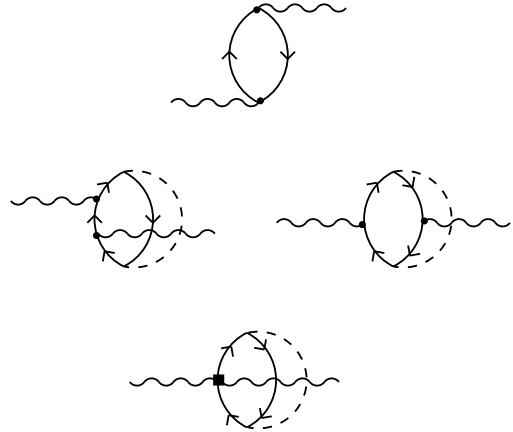}
}
\caption{Other diagrams contributing to $\rho$ selfenergy in the nuclear medium.}
\label{fig:rho2}       
\end{figure}

\begin{figure}
\resizebox{0.45\textwidth}{!}{%
  \includegraphics{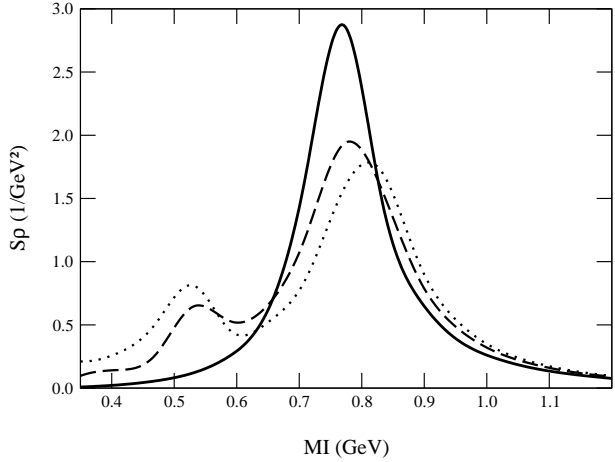}
}
\caption{ $\rho$ spectral function in vacuum: solid line, for $\rho=\rho_0/2$: dashed line and for $\rho=\rho_0$: dotted line.}
\label{fig:rho3}       
\end{figure}

The $\rho$ meson decays in vacuum into two pions. The large modification of the pion properties in nuclei discussed in the previous section implies changes of the $\rho$ properties, i.e., in the medium new decay channels are open because of the coupling of pions to baryon-hole excitations, see Fig. \ref{fig:rho1}. Other new decay channels appear because  of the direct coupling of the $\rho$ to the nucleon, sometimes producing some baryon-hole excitation, as shown in  Fig. \ref{fig:rho2}. In particular,
the excitation of the $N^*(1520)$ is very relevant. The contribution of these and other pieces is done using standard many body techniques. 
Details can be found in Refs. \cite{Cabrera:2000ct,Cabrera:2000dx}.
The obtained spectral function is depicted in Fig. \ref{fig:rho3}. Although some strength 
appears at low masses, due to the coupling to  $N^*(1520)$, and there is some 
widening,
the $\rho$ peak moves very little toward a mass higher than in vacuum. This is in contrast with
popular dropping mass scenarios. Several microscopic calculations have obtained very similar results \cite{Urban:1998eg,Post:2000qi} which seem to have been confirmed by recent experimental data \cite{Damjanovic:2005ni} when compared with various theoretical approaches \cite{vanHees:2006iv}.

\begin{figure}
\begin{center}
\resizebox{0.40\textwidth}{!}{%
  \includegraphics{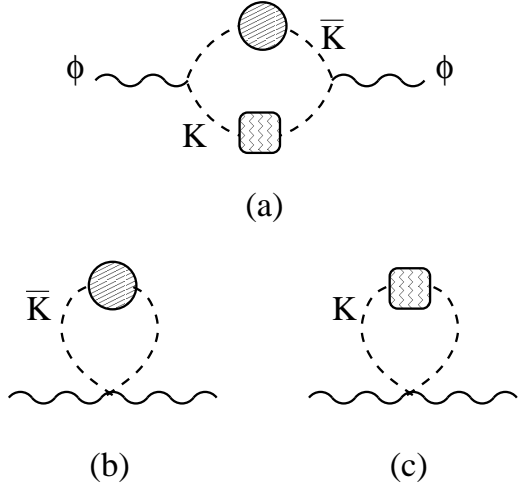}
}
\caption{ Diagrams contributing to $\phi$ selfenergy.}
\label{fig:phi1}       
\end{center}
\end{figure}

\begin{figure}
\begin{center}
\resizebox{0.40\textwidth}{!}{%
  \includegraphics{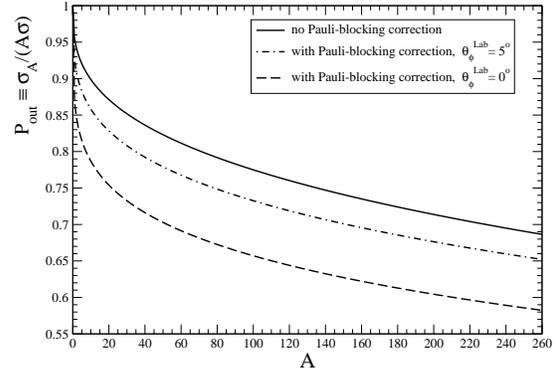}
}
\caption{Ratio of Nucleus to $A$ times nucleon cross sections, $\sigma_A/(A\,\sigma$, for  $\phi$ production as a function of A ($p_{\phi}=2 GeV$) }
\label{fig:phi2}       
\end{center}
\end{figure}

The $\phi$ meson in medium properties have been studied along similar lines in Refs. \cite{Oset:2000eg,Cabrera:2002hc}.
The $\phi$ decays primarily in $K\bar{K}$ and in the medium  the kaons have an spectral function quite different  
from vacuum, as discussed previously. The consideration of these effects, depicted in the diagrams shown in Fig. \ref{fig:phi1},
 leads to a width one order of magnitude larger than in
vacuum and also to some small mass reduction. In \cite{Cabrera:2003wb}, the loss of flux in photoproduction reactions 
was proposed as a possible experimental test of the model, see Fig. \ref{fig:phi2}. 
Later experiments  
\cite{Ahn:2004id} have confirmed the importance of medium effects
suggesting a $\phi$ width even larger than expected that might indicate a very large 
$\phi N$ cross section \cite{Muhlich:2005kf}. The small reduction of the mass has also been confirmed recently
\cite{Muto:2005za} although a detailed comparison of the theoretical models with data is still to be done. 

The theoretical study of the $\omega$ meson in the  medium is technically more complex due to its strong coupling to three-body channels although its narrow width makes it very well suited for experimental observation. 
Recently, the CBELSA/TAPS Collaboration has published some interesting results showing medium effects on the 
$\omega$ meson \cite{Trnka:2005ey}. A possible explanation of the data is the reduction of the $\omega$ mass. Other possibilities related to final state interactions of the detected $\pi^0$ are currently been studied.

\section{Scalar mesons ($\sigma$, $a_{0}$, $f_{0}$ and $\kappa$)}
\label{sec:sca}
\begin{figure}
\resizebox{0.45\textwidth}{!}{%
  \includegraphics{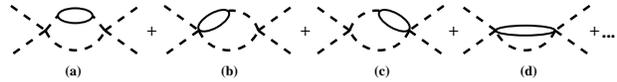}
}
\caption{Diagrams contributing to meson meson scattering in the nuclear medium.}
\label{fig:sig1}       
\end{figure}

One of the major successes of chiral unitary models is the description of meson-meson scattering amplitudes  up to energies above 1 GeV. Some resonances appear that can be identified with the $a_{0}$ and $f_{0}$ mesons. Much further from the real axis appear the poles corresponding to the $\sigma$  and $\kappa$ mesons. The medium modifications of these resonant states can be obtained calculating meson meson scattering but now including the proper in medium selfenergies of the mesons as depicted in Fig. \ref{fig:sig1}, apart from some vertex corrections that produce only minor effects.

\begin{figure}
\resizebox{0.45\textwidth}{!}{%
  \includegraphics{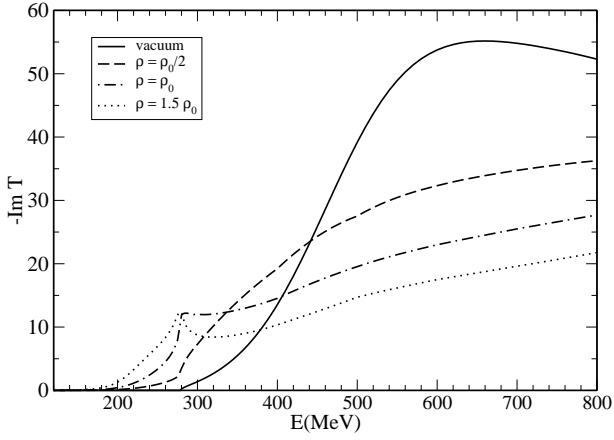}
}
\caption{Imaginary part of the pion pion scattering amplitude in the $\sigma$ channel at several densities.}
\label{fig:sig2}       
\end{figure}

The properties of the  
$\sigma$ meson in medium are of particular interest because of theoretical predictions claiming that at normal nuclear densities it could become much lighter and narrower, due to chiral symmetry restoration, and also because of some experimental results \cite{Camerini:1993ac,Messchendorp:2002au} showing an
enhancement of two pion production in the isoscalar channel at very low invariant masses that might be due to the 
in medium $\sigma$ meson. Indeed, in chiral unitary models the $\sigma$ spectral function has a large enhancement at low masses \cite{Chiang:1997di}, see Fig. \ref{fig:sig2} where the imaginary part of the $\pi\pi$ 
scattering amplitude is shown. 
This enhancement is related in these models to the position  of the  $\sigma$ pole 
that at $\rho=\rho_0$ appears at low mass and closer to the real axis \cite{Cabrera:2005wz}.
The calculation of two pion photoproduction making use of these techniques for the nuclear effects and starting from a good microscopical model for the reaction in vacuum, has found good agreement with experiment
\cite{Roca:2002vd} for both total and differential cross sections.
\begin{figure}
\resizebox{0.45\textwidth}{!}{%
  \includegraphics[angle=270]{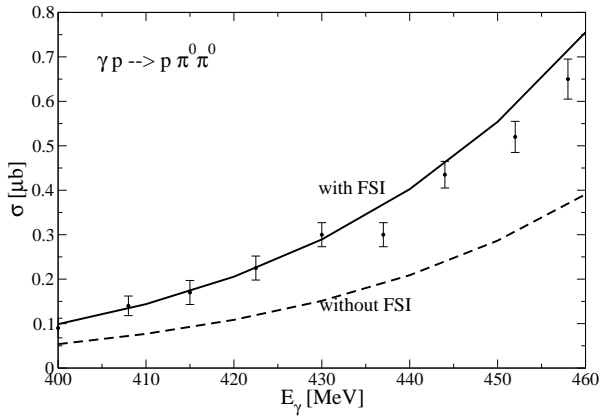}
}
\caption{Cross section for 2 $\pi^0$ photoproduction in nuclei as a function 
of the energy with and without final state interaction between the two pions 
in the I=0 channel. Experimental points from Ref. \cite{wolf}.}
\label{fig:sig3}       
\end{figure}

However, the presence of additional medium effects makes difficult the isolation of possible $\sigma$ effects \cite{Muhlich:2004zj,Alvarez-Ruso:2005dh}, which are small due to the low effective 
density of these processes. Nonetheless, these results together with those 
studying pion induced two pion production \cite{Rapp:1998fx,VicenteVacas:1999xx} support the possibility of testing the calculations when experiments with better statistics and for heavier nuclei are carried out.

Also interesting is the case of the $\kappa$ and $\bar{\kappa}$ that because 
of the different interaction of kaons and antikaons with the medium  get a 
different mass\cite{Cabrera:2004kt}. The large attraction that the nucleus
exerts over the pions makes these
resonances lighter and also narrower.
However, they do not become as narrow as to be clearly visible experimentally. 
Nonetheless, these results are very sensitive to the strength of the antikaon potential so controversial nowadays. If very deep antikaon potentials were confirmed that would automatically imply the existence of a low mass narrow $\bar{\kappa}$ in nuclei. In addition to the $\kappa$ pole positions, also the $K\pi$  and $\bar{K}\pi$ s-wave  scattering amplitudes at finite densities were studied.
Strong modifications were found for both channels that also became different although they are equal in vacuum.
The most noticeable effect is a peak found at low invariant masses of the $K\pi$ system that could be observed experimentally in reactions where a s-wave low energy $K\pi$ pair is produced.

Other dynamically generated mesons like the $a_{0}$ and $f_{0}$ have also been studied theoretically using the same method in Ref. \cite{Oset:2000ev}. These two resonances are relatively narrow and couple strongly to the $K\bar{K}$ channel. A large growth of their widths was obtained but without relevant changes in their masses.
\begin{figure}
\resizebox{0.45\textwidth}{!}{%
  \includegraphics{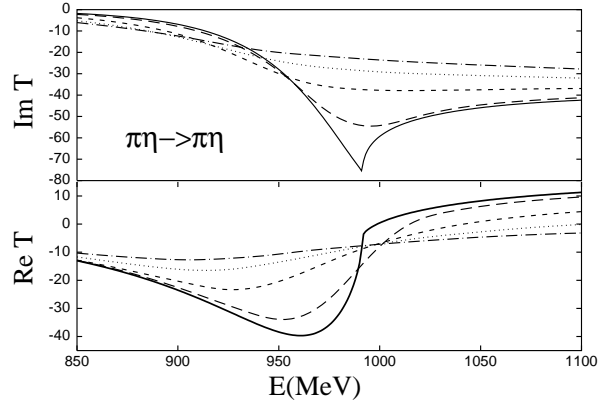}
}
\caption{Real Imaginary part of the $\pi\eta$ scattering amplitude in the I=1 channel at several densities. Solid line, free amplitude; long dashed line, 
$\rho=\rho_0/8$; short dashed line, $\rho=\rho_0/2$; dotted line, 
$\rho=\rho_0$; dashed dotted line, $\rho=1.5\rho_0$.}
\label{fig:eta}       
\end{figure}
As an example, we show in Fig. \ref{fig:eta} the $\pi\eta$ scattering amplitude. Whereas in vacuum there is a peak, related to the $a_{0}(980)$ pole, in medium, even at low densities the peak practically disappears.

\section{Baryons}
\label{sec:bar}

\begin{figure}
\resizebox{0.45\textwidth}{!}{%
  \includegraphics{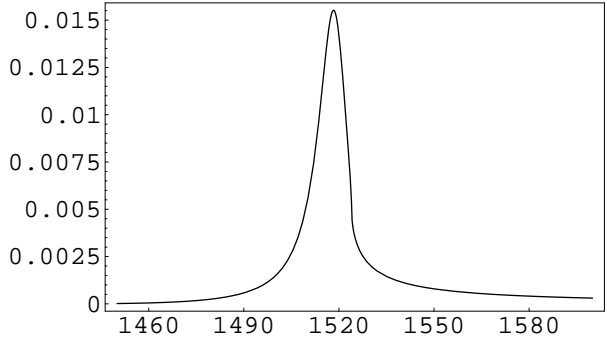}
}
\caption{ $|T_{\bar{K}N\rightarrow\pi\Sigma^*}|^2$ in units of MeV$^{-2}$ showing the $\Lambda(1520)$ peak.}
\label{fig:lam}       
\end{figure}

We will discuss first the case of the
$\Lambda(1520)$  which is generated from the $s$-wave interaction of the decuplet of $3/2^+$ baryons
with the octet of pseudoscalar mesons\cite{Kolomeitsev:2003kt,Sarkar:2004jh}. 
For a better description, the inclusion of the $KN$ and $\pi\Sigma$ $d$-wave channels was also required to get a good agreement on the position of the peak and the width of the resonance\cite{Sarkar:2005ap,Roca:2006sz}, see Fig. \ref{fig:lam}.

Its relatively small width makes this resonance suitable for the experimental verification  of the predicted medium effects.

In vacuum,  $\Lambda(1520)$ couples strongly to the $\pi \Sigma^*(1385)$ channel, even when the decay
is largely suppressed because of the small phase space. However, in the nuclear medium, this decay channel
is much enhanced due to both the pion and  $\Sigma^*(1385)$ selfenergies, leading to a 
predicted width $\sim 5$ times larger than in vacuum at normal nuclear matter densities \cite{Kaskulov:2005uw}. 
This large effect could be observable experimentally, studying the A-dependence of $\Lambda(1520)$ production reactions
\cite{Kaskulov:2006nm}.

As mentioned above, the antikaon in medium selfenergy was calculated in a  self-consistent way in Ref. \cite{Tolos:2006bj}. This also implies the study of medium effects of several hyperons relevant for the calculation. Their results show a downwards shift for the $\Lambda$ and $\Sigma$ masses of -30 MeV at normal nuclear density and a large enhancement of
 the $\Sigma^*$ width. The dynamically  generated $\Lambda(1405)$ is strongly diluted at $\rho=\rho_0$ and remains only as a small bump close to its vacuum position.
 
\section{In medium baryon baryon interaction}
\label{sec:inm}
\begin{figure}
\begin{center}
\resizebox{0.45\textwidth}{!}{%
  \includegraphics{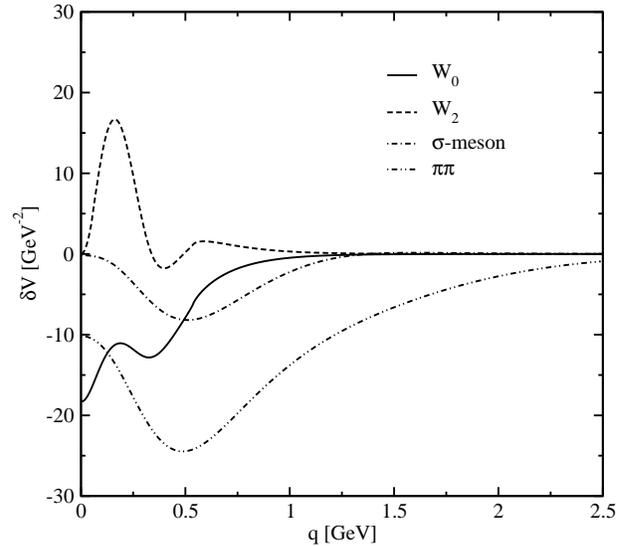}
}
\end{center}
\caption{Modification of the $NN$ interaction at normal nuclear matter density
in momentum space.
Solid line: central part of OPEP, dashed line: tensor part of  OPEP, 
dot-dashed: correlated $\sigma$-meson exchange and dot-dot-dashed curve:
uncorrelated $\pi \pi$ exchange.}
\label{fig:murat}       
\end{figure}

The scalar isoscalar channel is a very relevant part of the nucleon nucleon 
interaction, basically mediated by the exchange of two correlated ("$\sigma$") 
or uncorrelated pions. The $\sigma$ part of the channel is strongly modified 
in the nuclear medium, as shown in a previous section, and this could lead to 
some effects on the $NN$ interaction in the medium. This topic was analyzed 
in Ref. \cite{Kaskulov:2005kr}. The corrections for the $NN$ interaction in 
the medium were found to be sizable, see Fig. \ref{fig:murat}. 
We also found that  short range correlations are fundamental to produce only 
moderate changes. Further studies are still required to analyze the stability 
and properties of nuclear matter with potentials incorporating these results.

The central part of the $\Lambda N$ and $\Lambda \Lambda$ potential has also 
been studied including the correlated,
which in this case involves both $\sigma$ and $\kappa$, and uncorrelated 
two-meson exchange\cite{Sasaki:2006cx}. We find a short range repulsion 
generated by the correlated two-meson potential which also produces an 
attraction in the intermediate distances region. This interesting feature 
cannot be reproduced by the exchange of a $\sigma$ meson.
The uncorrelated two-meson exchange produces a sizable attraction in all cases. 

\section{Conclusions}
\label{sec:con}
In this work we have presented some recent results on the  study of hadron
properties. The use of $\chi$PT to incorporate low energy constraints on the
scattering amplitudes  and of coupled channels unitary models which generate
dynamically meson and baryon resonances provides a suitable framework to study
nuclear medium effects  on  those generated resonances. These medium effects are
considered using  standard techniques of many body Quantum Field Theory.
In this approach, we have shown results for several scalar mesons: $\sigma$, 
$\kappa$, $a_0(980)$ and $f_0(980)$ and for some baryonic resonances like 
$\Lambda(1520)$ and   $\Lambda(1405)$ and also for the controversial antikaon
optical potential. Some of the results presented here already have good
experimental support and others will soon be tested.

We have also discussed the scalar isoscalar part of the nucleon
nucleon  interaction and how this piece, partly mediated by the $\sigma$, is
strongly  modified in a nuclear medium. It must be considered that in our
approach the  $\sigma$ is only a resonant two pions state and the interaction
resulting from the exchange of this resonance cannot be fitted by a single
meson exchange. 

Finally, 
the pion and kaon in medium selfenergies have important consequences for the
decay of vector mesons that have also been analysed with our methods providing
an alternative   picture to other calculations. We find quite small mass changes,
consistent with many experimental data, but large width enhancements and even
complicated spectral functions with several bumps produced by the strong
coupling of the vector mesons to some particular excitation mode. This shows the
need of detailed microscopic calculations that properly take into account
the relevant degrees  of freedom in the nuclear medium.

\section{Acknowledgments}
\label{sec:ack}
This work is partly supported by the Spanish CSIC and JSPS collaboration, the
DGICYT contract number BFM 2003-00856,
and the E.U. EURIDICE network contract no. HPRN - CT -2002-00311. 
This research is part of the EU Integrated Infrastructure Initiative
Hadron Physics Project under contract number RII3 - CT -2004-506078.

%
%

\end{document}